# Quantification of geogrid lateral restraint using transparent sand and deep learning-based image segmentation

David H. Marx[1,a], Krishna Kumar[1,b] and Jorge G. Zornberg[1,c]

[1] Department of Civil Architectural and Environmental Engineering, The University of Texas at Austin, United States of America.

[a] dawie.marx@utexas.edu (corresponding author), [b] krishnak@utexas.edu, [c] zornberg@mail.utexas.edu

## Abstract

An experimental technique is presented to quantify the lateral restraint provided by a geogrid embedded in granular soil at the particle level. Repeated load triaxial tests were done on transparent sand specimens with geosynthetic inclusions simulating geogrids. Particle outlines on laser illuminated planes through the specimens were segmented using a deep learning-based segmentation algorithm. The particle outlines were characterized in terms of Fourier shape descriptors and tracked across sequentially captured images. The accuracy of the particle displacement measurements was validated against Digital Image Correlation (DIC) measurements. In addition, the method's resolution and repeatability is presented. Based on the measured particle displacements and rotations, a state boundary line between probable and improbable particle motions was identified for each test. The size of the zone of probable motions could be used to quantify the lateral restraint provided by the inclusions. Overall, the tests results revealed that the geosynthetic inclusions restricted both particle displacements and rotations. However, the particle displacements were found to be restrained more significantly than the rotations. Finally, a unique relationship was found between the magnitude of the permanent strains of the specimens and the size of the zone of probable motions.

## 1. Introduction

Geogrids have been successfully used to improve several aspects of roadway performance. Applications of geogrids in roadways include the stabilization of roadways founded on soft ground, effectively increasing the subgrade's bearing capacity, and can also stabilization of unbound aggregates, thereby extending pavement life or allowing for the use of reduced pavement cross sections (Beranek 2003; Zornberg 2017). In applications involving stabilization of soft subgrades the tensioned membrane and vertical restraint mechanisms dominate due to the significant roadway deformations occurring. In contrast, for applications involving stabilization of unbound aggregates, lateral restraint of the particles is the dominant mechanism. For these applications the geogrids stiffens the unbound aggregate layer by providing lateral restraint to the aggregate matrix through soil-geogrid interaction (Beranek 2003; Zornberg 2017). When used for stabilization of unbound aggregates (e.g. for base course reduction), the geogrids are typically placed within the base layer or between the base and subsurface layers of the roadway. The increased stiffness of the unbound aggregate layer results in reduced rutting of the road, as multiple studies have demonstrated (e.g. Haas et al. 1988; Webster 1993; Collin et al. 1996; Moghaddas-Nejad and Small 1996).

The increased stiffness of the stabilized aggregate layer has been measured in both full scale studies and laboratory testing (e.g. Abu-Farsakh et al. 2007; Kwon and Tutumluer 2009; Nair and Latha 2015; Byun et al. 2019; Yu et al. 2019; Kang et al. 2022). However, limited studies have been done at the particle level on quantifying the mechanism of lateral restraint of the soil matrix due to interlock with a geogrid. Konietzky et al. (2004) and McDowell et al. (2006) used Discrete Element Modelling (DEM) to simulate the particle level behaviour of triaxial specimens with geosynthetic inclusions under cyclic loading. Similar experimental evaluation would require particle level measurements in order to capture the lateral restraint mechanism due to soil-geogrid interaction. While these type of experimental measurements of soil-geogrid interaction have been conducted in the past, the focus of these previous studies has mostly been on reinforcement applications. These studies have investigated the impact of geogrids on strength under comparatively large displacements and monotonic loading, rather than assessing the influence of geogrids on stiffness under low displacements and cyclic loading (Dyer 1985; Zhou et al. 2012; Abdi and Mirzaeifar 2017; Peng 2017; Chen et al. 2021; Derksen et al. 2021). Consequently, although there is consensus in literature that lateral restraint is the mechanism governing the improvement provided by geogrids in stabilization of unbound aggregate applications, actual evaluation of such mechanism has remained, at best, unquantified.

The work presented in this paper involves an experimental technique developed to quantify the lateral restraint of particles in triaxial specimens under repeated loading by geosynthetic inclusions simulating geogrids. The technique entails: 1) the use of transparent sand with laser illumination to visualize the motions (displacements and rotations) of individual particles, 2) the use of a deep learning-based segmentation method to segment images of the transparent sand into individual particles, 3) tracking of the segmented particles by matching their Fourier shape descriptors, and 4) a comparison of particle motions for triaxial specimens with and without the inclusions.

## 2. Experimental program

### 2.1. Testing setup

In conventional geotechnical testing, a triaxial specimen is a unit cell in which the stress state is expectedly uniform and well defined. Yet, this is not the case for a triaxial specimen with a geogrid inclusion, as the stress state across the inclusion is discontinuous. Nevertheless, a triaxial test setup still allows for soil-geogrid interaction to be studied under controlled boundary conditions. In this study, repeated load triaxial testing on 50 mm-diameter triaxial specimens were used to model the development of the lateral restraint provided by geogrids. A brief overview of the experimental setup adopted for the tests conducted in this study is presented below. Details of the setup are discussed in Marx and Zornberg (2022).

The experimental setup for this study consisted of a conventional triaxial cell and load frame placed inside a custom imaging frame. Key elements of the imaging frame, such as the image alignment markers, cameras, and lasers, are indicated in Figure 1. The two lasers shown in the figure were placed at 60° relative to each other to illuminate 2D planes through the centre of the specimens (see Section 2.2). The power of lasers used in this study were 350 mW and 450 mW respectively, both with a wavelength of 638 nm. In addition, two Canon EOS 5DS R cameras fitted with Sigma 50 mm f/1.4 lenses were placed at 90° relative to each laser to capture images of the laser-illuminated planes through the triaxial specimens.

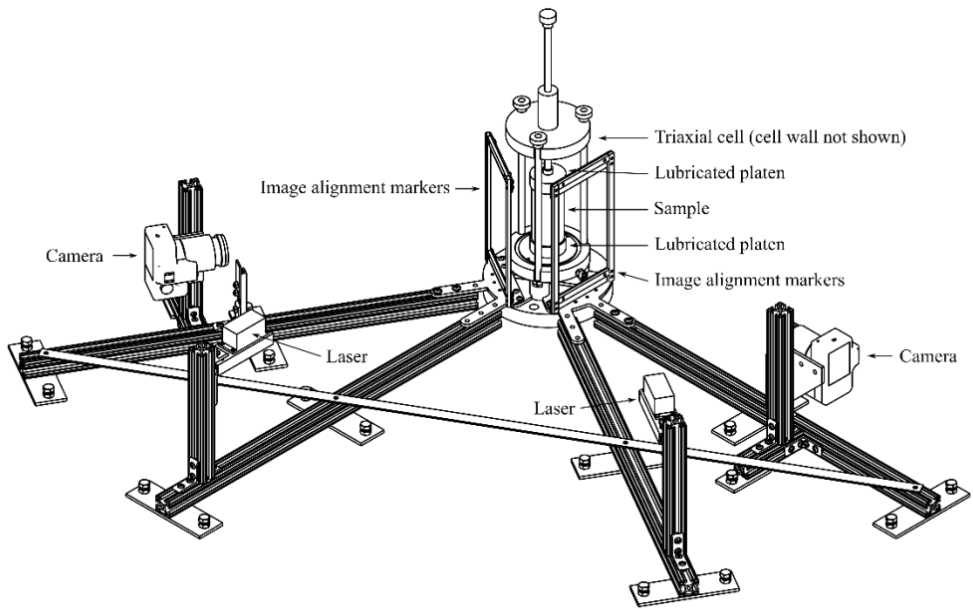

Figure 1. Experimental setup showing the position of imaging frame, triaxial cell and specimen.

Despite clamping the cameras to the imaging frame, minute movement of the cameras may occur during testing. Thus, the images of the specimens had to be aligned relative to a static reference marker. Red Light Emitting Diodes (LEDs) fastened to a frame were used as alignment markers, following the approach adopted by Stanier et al. (2012). The alignment markers were placed in a rectangular border around the specimen.

In a conventional triaxial test the porous stones or filter paper can provide lateral restraint to the adjacent soil, resulting in so-called 'dead zones' (Head 2014). For specimens with an aspect ratio smaller than 2:1 the two 'dead zones' can overlap, resulting in an artificially strengthened specimen (Bishop and Green 1965). Similarly, for specimens with geogrid inclusions, the 'dead zones' may overlap with the soil body restrained by the geogrid. Consequently, it may be hard to isolate the contribution provided by the geogrid from the effect imposed by the boundary conditions. In these tests enlarged, lubricated platens with drainage holes were used to minimize the lateral restraint at the ends of the specimens. Two layers of black silicone rubber, lubricated with thin layers of vacuum grease, were used for each platen.

## 2.2. Transparent sand

The triaxial tests with geosynthetic inclusions were done on transparent sand to allow for individual particles to be visualized. Transparent sand can be manufactured by index matching fused quartz with mineral oil (Ezzein and Bathurst 2011). Due to the matching refractive indices, light transmits through the specimen, rather than reflecting at the particle surfaces. Thus, the specimen is transparent. The depth of transparency is influenced by the presence of air bubbles and contaminants, the quality of the refractive index match, and the number of particles between the observer and the section of interest (Black and Take 2015; Hunter and Bowman 2018).

Fused quartz is a form of quartz sand (silicone dioxide) that has been melted and cooled to form a glass (non-crystalline) (Ezzein and Bathurst 2011). A photo of the fused quartz used in this study is shown in Figure 2a. The material passed the #6 sieve (3.35 mm) and was retrained on the #8 sieve (2.36 mm). Consequently, it corresponded to a uniformly graded sand with a $D_{50}$ of 2.9 mm. The minimum and maximum void ratios were 0.72 and 0.85, according to (ASTM D4253; ASTM D4254).

The median particle roundness was 0.420 and the median area sphericity was 0.560. The area sphericity was defined as the ratio of the length of the particle to the width, following Zheng and Hryciw (2015). Roundness is representative of the sharpness of the particle corners, with a maximum of 1.0 for a circle (Wadell 1932). Both the roundness and area sphericity were calculated from 2D images of the particles using the software developed by Zheng and Hryciw (2015). Thus, particles are highly angular and flaky. Therefore, while the smallest particle dimension is 2.9 mm, the largest can measure up to 5 mm in diameter

The fused quartz was saturated with a mixture of two oils: 52% Puretol 7 Special and 48% Paraflex HT4, manufactured by Petro-Canada (Peng and Zornberg 2019). A photo of a partially saturated sample of fused quartz is shown in Figure 2b.

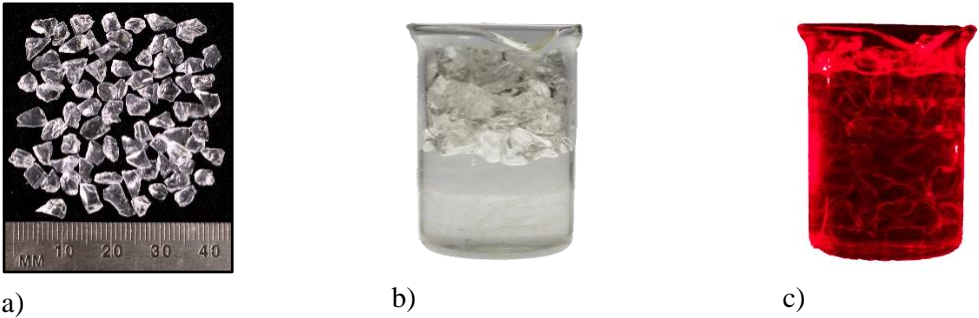

a)                              b)                              c)

Figure 2. Transparent sand: a) fused quartz particles, b) fused quartz particles partially saturated with mineral oil, and c) laser illuminated plane through saturated fused quartz.

A perfect match of the refractive indices is particularly difficult to achieve. The inevitable minor mismatch can be exploited by illuminating a plane through the sample using a thin, sheet laser. The laser refracts at the edge of each particle it intersects. The refraction brightens the particle edges, allowing for the outlines of individual particles along the 2D plane to be identified as visible in Figure 3c (Sadek et al. 2003; Peng and Zornberg 2019; Derksen et al. 2021).

The clarity of the particle outlines on a laser-illuminated plane is affected by the same parameters that affects transparency. In this study, the clarity of the particle outlines was maximized by minimizing the number of particles between the laser plane and the cameras. Thus, the use of a 50 mm specimen and a uniformly graded quartz.

### 2.3. Geosynthetic Inclusions

The stresses applied to a triaxial specimen are axisymmetric. Thus, a geogrid used to stiffen the triaxial specimen should ideally provide axisymmetric lateral restraint. The simplest geometry that provides axisymmetric lateral restraint is an annular inclusion, as shown in Figure 4a. However, when a thin annular ring is used to stiffen a triaxial specimen, only a limited volume of soil will interact with the geogrid. Two concentric rings could instead be used to increase the lateral restraint; however, this would lead to a greater restraint in the radial direction than in the circumferential direction. Probably the only geometric pattern that would provide equivalent lateral restraint in all directions, other than a ring, is a honeycomb pattern as shown in Figure 4b.

The two types of honeycomb inclusions, shown in Figure 4c and d, were used in this study to simulate the lateral restraint provided by a geogrid. These geosynthetic inclusions were cut from a 3 mm-thick polypropylene sheet using a water-jet cutter. The thickness and stiffness of the geosynthetic inclusions were selected to minimize deformation and potential breakage of the ribs during testing, as lateral restraint was the focus of this study. Both geosynthetic inclusions share the same rib thickness, but differ in aperture size. For Honeycomb inclusion 1 (H1) the ratio of aperture width to particle size ($A/D_{50}$) was 4.0 and for Honeycomb inclusion 2 (H2) it was 2.2.

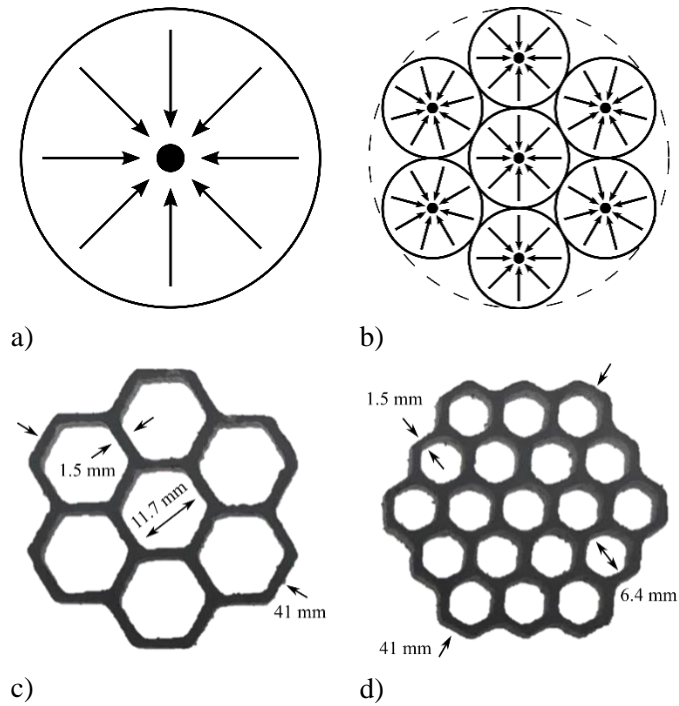

Figure 4. Geosynthetic inclusions embedded in triaxial specimens: a) lateral restraint by an annular inclusion, b) lateral restraint by honeycomb inclusion, c) inclusion H1 and d) inclusion H2.

## 2.4. Testing procedure

The purpose of this study was to investigate the lateral restraint provided by geogrids in stabilization applications of unbound aggregates, rather than reinforcement applications. In stabilization applications of roadways, design is governed by the serviceability criteria (i.e. deformation). Consequently, the relevant strain levels are significantly smaller than the comparatively large strains required to reach the ultimate strength of a geogrid. To simulate these loading conditions the triaxial specimens were tested with emphasis on displacements at working stress conditions (rather than ultimate conditions) and under repeated loading (rather than monotonic loading). Furthermore, the effectiveness of the geosynthetic inclusions was measured in terms of the permanent strain that accumulated, rather than the ultimate strength of the specimens. The triaxial samples stiffened with geosynthetic inclusions, measuring 50 mm in diameter and 100 mm in height, were constructed inside a custom, transparent silicone membrane (Marx and Zornberg 2022).

A summary of the six triaxial specimens tested in this experimental program is presented in Table 1. Two control specimens (CM-25-1, CM-25-2) were sheared monotonically until failure. A confining stress of 25 kPa was selected as representative of the typical residual stresses in a road base/subbase layer after compaction (Uzan 1985; Selig 1987; Barksdale et al. 1997). The results of these control tests were used to assess the repeatability of the setup and to determine the load levels used for the repeated load testing.

The repeated loading tests consisted of four stages, as summarised in Table 2. The load applied in each stage was defined based on the stress-strain response of the monotonic tests (presented in Section 4.4). The initial stage of the tests (Stage 1) involved loading the specimens within the linear-elastic regime. In testing Stage 2, the load was intended to cycle into the non-linear elastic regime for the control specimens. In testing Stage 3, the load was intended to cycle into the start of plastic deformation. Finally, in testing Stage 4 the specimens were monotonically sheared up to the shear strength of the control specimens (140 kPa). The number of cycles adopted in the testing program was selected to account for the slow loading rate as well as the large volume of image data that was generated for test. The top platen was uncoupled from the loading ram to avoid the development of tension in the triaxial specimen during unloading

Four specimens were tested under repeated loading: one control (CR), one stiffened with H1 (H1-25), one stiffened with H2 (H2-25) and one stiffened with H1 tested at a higher confining stress of 50 kPa (H1-50). The geosynthetic inclusions were placed horizontally, in the centre of the specimens. An image captured of a laser-illuminated plane through specimen H1-25 before loading, is shown in Figure 5.

Table 1. Summary of triaxial specimens tested as part of this study

| Specimen | Confining stress (kPa) | Geosynthetic inclusion | Loading pattern |
|---|---|---|---|
| CM-25-1 | 25 | None | Monotonic |
| CM-25-2 | 25 | None | Monotonic |
| CR-25 | 25 | None | Repeated |
| H1-25 | 25 | H1, centre | Repeated |
| H2-25 | 25 | H2, centre | Repeated |
| H1-50 | 50 | H1, centre | Repeated |

Table 2. Summary of loading stages in tests involving repeated loading

| Loading stage | Applied stress [kPa] | Equivalent in monotonic test | | Number of cycles |
|---|---|---|---|---|
| | | Strain [%] | Loading regime | |
| 1 | 70 | 0.5 | Linear-elastic | 5 |
| 2 | 85 | 1 | Non-linear elastic | 5 |
| 3 | 100 | 2 | Start of plastic | 3 |
| 4 | 140 | 7.5 | Failure | 1 |

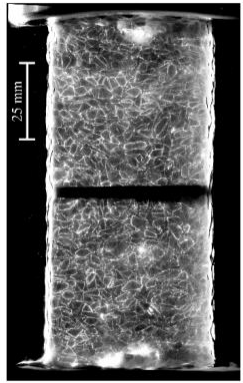

Figure 5. View of a laser-illuminated plane through specimen H1-25 before loading.

The six specimens prepared for this study were compacted in eight layers by tamping. The specimens had an average void ratio of 0.71. This comparatively high density was adopted in an attempt to minimize heterogeneity within the soil matrix of each triaxial specimen, as well as to facilitate repeatability of the target density across specimens. Consequently, any discrepancy in the measured particle motions between triaxial specimens CM-25-1 and CM-25-2 would be due to a deficiency of the new measurement technique, rather than due to issues related to specimen preparation. In addition, the density adopted for the specimens was consistent with the heavily compacted aggregates in a road base layer

A back pressure of 300 kPa was applied to facilitate dissolution of air bubbles that may be trapped between the particles and thus improve the transparency of the specimens. The B-value for all tests was in excess of 0.99. Oil was used both as the saturating liquid and the confining liquid.

The deviatoric stress was calculated considering the area corrections proposed by Head (2014) and the membrane corrections proposed by Duncan and Seed (1967). Volume reduction due to membrane penetrations was considered to be negligible (Hjortnaes-Pedersen and Molenkamp 1982; Head 2014).

Both the monotonic and the repeated load tests were sheared at a rate of 6% axial strain per hour. The comparatively slow rate of shearing adopted in this study was required to facilitate pore pressure dissipation through the small drainage holes in the lubricated platens (Head 2014). Furthermore, the sample had to be sheared slow enough such that no significant particle motion would occur while the camera shutter was open. A relatively low shutter speed of 0.6 s was required to achieve adequate exposure of the images.

Alternating images were captured during testing of the two laser-illuminated planes through the specimens. For the monotonic shear tests, images were captured at increments of 0.02% axial strain. During the repeated loading tests the frequency was increased to 0.01% axial strain due to the smaller range of applied strain.

## 3. Image processing

### 3.1. Pre-processing

The images of the specimens had to be pre-processed before particle motions could be measured. A combination of the OpenCV (Bradski 2000) and Scikit-Image (van der Walt et al. 2014) Python libraries was used for the pre-processing. A summary of the image processing pipeline is presented below. Additional details are provided by Marx and Zornberg (2022).

Colour (RGB) images of the planes were captured during testing. The colour images were predominantly monochromatic due to the red laser used for illumination (see Figure 2c). Thus, the images could be converted to greyscale without a significant loss of information.

The images were dominated by the bright intensities of the particle edges. To recover the detail in the darker regions of the images (e.g. particle interiors and poorly illuminated particle edges) the logarithm of the image greyscale values was calculated. This approach compressed the dynamic range, allowing for the darker regions to be represented with a larger number of grey values (bits) (Bovik 2009).

The single light source (the lasers) and heterogeneous refraction by the particles resulted in inconsistent exposure across the images. In order to achieve a uniform exposure, Contrast Limited Adaptive Histogram Equalization (CLAHE) was applied to the images (Zuiderveld 1994). CLAHE locally redistributes the pixel intensities across the range of available grey values (i.e. 256 for an 8-bit image).

The equalized images were then corrected for internal camera distortion as well as lens distortion. A discussion on camera and lens distortion can be found in White et al. (2003). Next, the corrected images were all aligned to the same coordinate system using the LEDs alignment markers as reference points.

The curved cell wall of the triaxial cell acts as a lens that results in horizontal distortion of light refracted from the particle edges. To correct the horizontal distortion of the images, the method presented by Zhang et al. (2015) was implemented.

Finally, the images were denoised with Gaussian smoothing (Bovik 2009). The result of the pre-processing pipeline was a series of undistorted, aligned images with clearly visible particles and a known pixel-to-millimetre calibration factor.

### 3.2. Image segmentation

#### 3.2.1. Analytical segmentation

The images collected for each triaxial test had to be segmented into individual particles in order to track particle displacements. However, analytical segmentation methods failed to properly segment the fused quartz from the pore space in the laser-illuminated planes. For example, Figure 6b shows the particles in Figure 6a segmented based on pixel intensity using the method described by Otsu (1979). Otsu's method failed as the particle edges, particle interiors and void space all shared the same intensities at different locations in the image. Consequently, multiple particles and void space were merged during the segmentation.

More advanced methods of analytical segmentation, such as *k*-means clustering (Szeliski 2010) also failed. In Figure 6c the segmentation of Figure 6a based on Simple Linear Iterative Clustering (SLIC) (Achanta et al. 2012) is shown. This technique groups clusters of perceptually similar pixels together into *k* classes. SLIC managed to segment the image into individual particles (Class 1). However, the accuracy of the segmentation was poor, as the edges of the particles were classified separate from the particles (Classes 2 and 3).

Rather than segmenting the particles and the void space, it was also attempted to first detect the particle outlines. These outlines are classified as "ridges", rather than "edges", in image analysis. A "ridge" occurs when the image intensity changes from dark to bright and back to dark (or the inverse). In contrast, at an edge the intensity changes from dark to bright and then remains constant. The ridges detected with the filter by Meijering et al. (2004) are shown in Figure 6d. Due to inconsistent lighting, surface roughness, and image noise, the particle outlines were discontinuous in the image and consequently also the response of the filter. Thus, ridge detection failed to separate the particles as well.

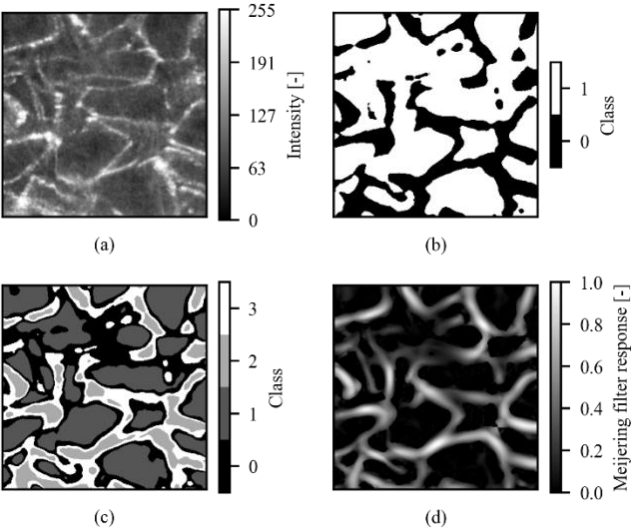

Figure 6. Analytical segmentation of images of transparent soil: a) raw image, b) segmentation with Otsu thresholding, c) segmentation with SLIC, d) Meijering edge filter.

### 3.2.2. Segmentation using the deep learning

Due to the poor performance of the analytical segmentation methods, the images were segmented with an algorithm based on deep learning segmentation, Cellpose (Stringer et al. 2021). Cellpose was originally developed to segment biological cells. Images of biological cells share several similarities with laser illuminated fused quartz. Similar to the fused quartz in Figure 7a, the human breast cancer cells in Figure 7b are defined by ridges, rather than edges. Fused quartz and the mouse microglia in Figure 7b consist of a matrix of multiple objects in contact with each other. The images of human ovarian cancer in Figure 7d suffers from inconsistent illumination inside the objects and a similar intensity of the void space and the interior of the objects.

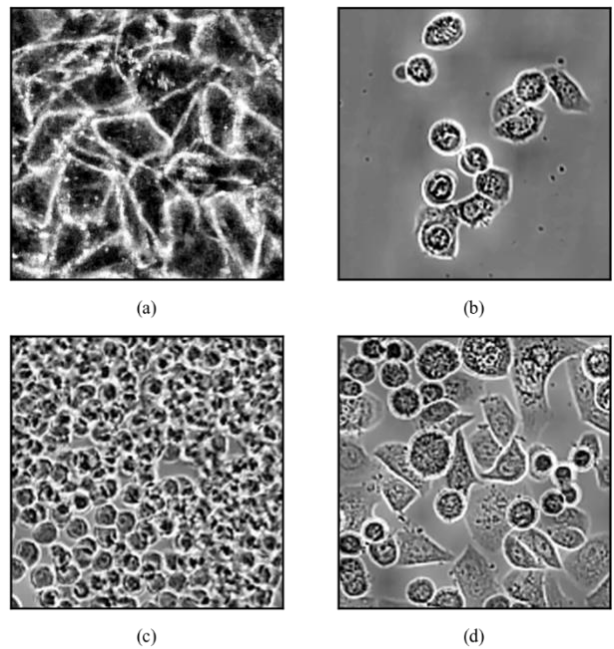

Figure 7. Similarity between images of fused quartz (a) and biological cells: b) human breast cancer cell, c) mouse microglia and d) human ovarian cancer (images from Edlund et al. (2021)).

The Cellpose algorithm circumvents the issues encountered with the analytical segmenting of images of fused quartz (or biological cells) by representing each pixel as a unit vector. The vectors are defined such that they each point to the next pixel (vector) that will eventually lead to the cell centre. Consequently, all vectors belonging to a given particle converges to the same point. A discussion on calculating the vector fields for already segmented images by simulating heat diffusion at the particle centres can be found in Stringer et al. (2021). An example of such a vector field for the fused quartz particles is shown in Figure 8b. The vectors and the outlines of the original particles are colour coded for clarity. Alternatively, each vector can be represented as vertical and horizontal gradient which leads to the next pixel in the chain to the particle centre. These gradients are shown in Figure 8c and Figure 8d.

The consequence of the vector representation of an image is that discontinuous particle outlines, image noise and global variations in intensity will not affect the definition of a particle. However, these vector representations cannot be calculated analytically if the particle outlines are not already known. Consequently, Cellpose uses a Convolutional Neural Network (CNN) to predict the vector fields. A neural network can be thought of as a composite function consisting of interconnected, non-linear functions. Further discussion of convolutional neural networks is presented in Appendix A1. A discussion of the architecture of the CNN used in Cellpose is presented in Appendix A2.

For each pixel in the image, the CNN predicts the vertical gradient (Figure 8c), the horizontal gradient (Figure 8d) and the probability that the pixel forms part of a particle. Subsequently, the gradients at all pixels with a probability greater than 0.5 are used to reconstruct the segmented particles. Cellpose outperformed general purpose segmentation CNNs such as Mask R-CNN (He et al. 2017) on images of cells.

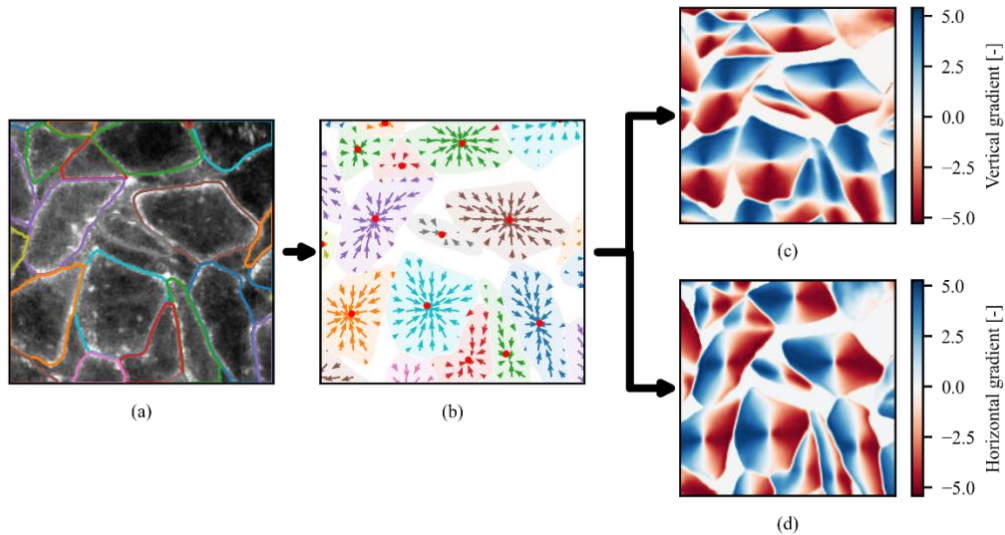

Figure 8. Demonstration of Cellpose's vector representation of an image a) segmented image, b) vector field, c) vertical gradient of vectors, d) horizontal gradient of vectors.

### 3.2.3. Training the Cellpose CNN to segment fused quartz

The CNN function coefficients (or weights) need to be calibrated for the network to provide the desired results. The process of 'calibrating' a neural network is known as training. Due to the non-linear nature of neural networks, training is an iterative process. A set of images with known particle positions is passed to the network multiple times. For each pass (epoch) the error of the prediction (loss) is calculated, and the network weights are updated accordingly.

Manually generating enough segmented fused quartz images to train the Cellpose CNN from the ground up was infeasible. Thus, an approach founded on transfer learning was implemented. Transfer learning is based on the principle that models that perform a similar function on different datasets will share some general characteristics (Pan and Yang 2010). Thus, a model developed to segment biological cells can be fine-tuned to segment fused quartz particles with a comparatively limited number of new training images.

The 'cyto2' model (Stringer and Pachitariu 2022) of Cellpose was used as the basis for the transfer learning. This model was initially trained on more than 600 manually segmented images of biological cells. To refine this model, six images of laser illuminated fused quartz were manually segmented using the Cellpose graphical user interface, yielding a total of 3,100 unique particles. These images were sampled from both control and stiffened specimens at the consolidation and shearing stages of the triaxial tests. The quality of the segmentation by Cellpose improved when the images were also denoised with total variation denoising (Getreuer 2012) in addition to the pre-processing discussed in Section 3.1. Total variation denoising was selected as it is an edge (and ridge) preserving denoising technique.

The six training images were augmented to increase the training set and thus the generalizability of the network (with regard to fused quartz). The augmentations entailed random rotation, flipping, cropping, resizing, and adjustment of the contrast and brightness of the input images. Data augmentation is based on the assumption that new information can be extracted from alternative views of the same input images (Shorten and Khoshgoftaar 2019). Twelve augmented images were generated from each of the original six images, yielding a total of 72 training images.

The augmented input data was divided into a training set (70%), validation set (20%) and testing set (10%) (Goodfellow et al. 2016). The images of the training set were used to refine the weights of 'cyto2' model for fused quartz. The error (loss) of the model was a function of the accuracy of the prediction of the vertical gradients, horizontal gradients and particle probability (see Equation A1 in Appendix A2). During training, the loss was also calculated for the validation set. However, these images were not considered when updating the weights. Training was stopped once the loss on the validation set converged. Convergence occurred after 750 epochs of training. Any further training would have resulted in overfitting the model to the training set.

The training and validation loss is calculated from the predicted horizontal and vertical gradients of the image, and not the actual accuracy of the particle segmentation. The accuracy $A$ of the segmentation can be calculated as (Stringer et al. 2021):

$$A = \frac{T_P}{T_P + F_N + F_P} \tag{1}$$

where $T_p$ are the true positives; $F_N$ are the false negatives; and $F_P$ are the false positives.

To determine whether a segmented particle is a true positive or a false positive the intersection over union (IoU) is calculated. The IoU is the ratio of the overlap between particle A and B to the combined footprint of A and B:

$$IoU = \frac{A \cap B}{A \cup B} \qquad (2)$$

An IoU value of 0.5 is commonly defined as a positive match. For an IoU of 0.5 two identical circles will overlap with 67%. For the 'cyto2' model, the initial segmentation accuracy on the test dataset of fused quartz images was 0.25 at an IoU 0.5. After refining the 'cyto2' model with the augmented images of fused quartz, the accuracy at an IoU of 0.5 increased to 0.69.

### 3.2.4. Validation of Cellpose segmentation methodology using Explainable AI

Deep neural networks consist of millions of trainable parameters (weights) and are effectively a "black box" for the user. Accordingly, the model user runs the risk that a model can make predictions for an image based on features that do not make physical sense. This risk is not negligible when applying a pre-trained model to a new dataset. Such as using a model trained to segment biological cells to segment images of fused quartz.

The discipline of "Explainable AI" attempts to improve the transparency and trustworthiness of neural networks without compromising their performance (Barredo Arrieta et al. 2020). The Score-CAM (Wang et al. 2020) metric was calculated to evaluate the explainability of the Cellpose network after it was trained on the images of fused quartz. The Score-CAM metric is a heatmap highlighting the areas in the image that contributed to the prediction by the network at a given pixel. Details of the Score-CAM implementation is presented in Appendix A3.

In Figure 9b, c, and d the Score-CAM heatmap for the three outputs of the Cellpose CNN is shown for the single pixel (◊) indicated in Figure 9a. The heatmap in Figure 9b shows that the prediction of the vertical gradient at the pixel is sensitive to the northern slope of the particle, the pore space, and the southern slope of the opposite particle. In addition, a few regions of the image unrelated to the specific pixel also contributed to the local prediction. This inconsistency is an artefact of the architecture of the Cellpose CNN that considers a significant portion of the image when making a prediction for a single pixel (see Appendix A2). In addition, some of this inconsistency can be attributed to boundary effects.

The Score-CAM heatmap in Figure 9c for the horizontal gradients follows the same trend as that for the vertical gradients. Finally, Figure 9d shows that the predicted probability that a pixel will fall inside a particle mostly depends on the edges of the surrounding particles. Thus, the heatmaps in Figure 9 verified that the predictions by the Cellpose model are grounded in the physical characteristics of the fused quartz images. Similar behaviour was observed when other pixels were evaluated. Consequently, it was deemed rational to use Cellpose to segment images of fused quartz.

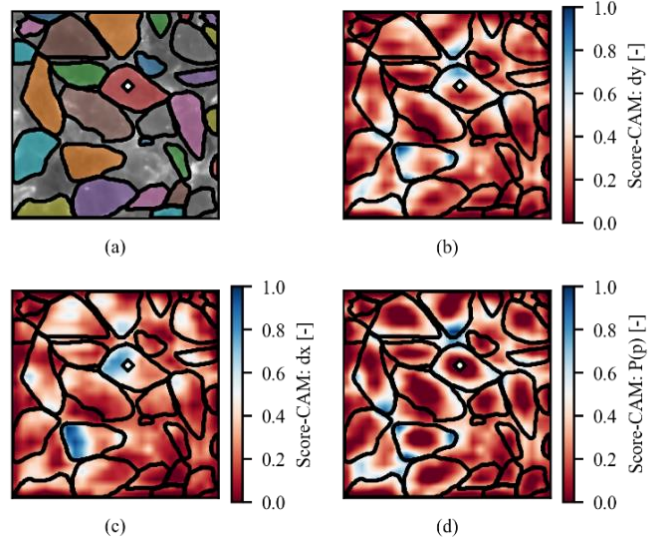

Figure 9. Score-CAM heatmaps for the indicated pixel: a) segmented image; b) Score-CAM for the vertical gradients; c) Score-CAM for the horizontal gradients; and d) Score-CAM particle probabilities.

## 4. Particle tracking and displacement measurement

### 4.1. Fourier Shape Descriptors

Inconsistent lightning and out-of-plane particle rotation may result in minor differences in the particle outlines between consecutive images. To effectively track these fluctuating particles, global representations of particle shape, known as Fourier descriptors (Burger and Burge 2013), were adopted in this study.

A Fourier series represents a periodic function as an infinite sum of sine and cosine functions. To calculate the Fourier descriptors of a particle outline, discrete coordinates $\boldsymbol{v_k} = (x_k, y_k)$ were sampled regularly along a particle's perimeter. These coordinates were then interpreted as points on the complex plane: $\boldsymbol{g_k} = x_k + i \cdot y_k$. Subsequently, the forward discrete Fourier transform of the complex plane is calculated (Burger and Burge 2013):

$$G_m = \frac{1}{M} \cdot \sum_{k=0}^{M-1} [x_k + i \cdot y_k] \cdot \left[ \cos\left(\omega_m \frac{k}{M}\right) - i \cdot \sin\left(\omega_m \frac{k}{M}\right) \right] \tag{3}$$

where $G_m$ is the Fourier shape descriptor for frequency $m$; $M$ is the signal length; and $\omega_m = 2\pi m$.

For the present study a variant of the Fourier shape descriptor calculation developed for polygons with arbitrarily spaced vertices was used (Burger and Burge 2013).

In Figure 10, a shape is displayed along with iterations of it reconstructed with an increasing number of Fourier descriptor pairs. The centroid of the shape, $G_0$ is also indicated. For only one pair of descriptors ($\langle G_{-1}, G_{+1} \rangle$), the reconstructed shape is the fundamental ellipse. As the number of descriptor pairs increases, the reconstructed shape represents the original shape in more detail. Thus, when calculating the Fourier descriptors of a fused quartz particle, the higher order coefficients will represent minor variations in the outline of a particle between two images. By filtering out these higher order coefficients, it is possible to match a particle between images despite fluctuations in its outline. In this study, nine Fourier descriptor pairs were established to be the optimal balance that allowed representing the essence of a particle without capturing too much detail.

Fourier analysis assumes that a signal is periodic, and closed particle boundaries are implicitly periodic. However, the Fourier descriptors may vary depending on where coordinates are sampled along the curve. Consequently, before matching the descriptors, they must be made invariant to the starting point. In addition, the descriptors were also made invariant to scale and rotation (Burger and Burge 2013).

Complex (phase-preserving) matching was used to calculate the distance between the descriptors $\boldsymbol{G}_1$ and $\boldsymbol{G}_2$ of two given particles ($FD_{dist}$):

$$FD_{dist}(\boldsymbol{G}_1, \boldsymbol{G}_2) = \left( \sum_{m=-M_p, m \neq 0}^{M_p} |\boldsymbol{G}_1(m) - \boldsymbol{G}_2(m)|^2 \right)^{\frac{1}{2}} \tag{4}$$

where $M_p$ is the number of Fourier descriptor pairs.

The centroid ($G_0$) was ignored during matching to make the coefficients invariant to translation.

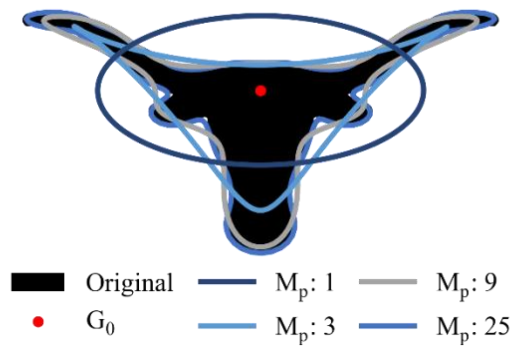

**Figure 10.** Shape reconstruction for an increasing number of Fourier shape descriptor pairs.

### 4.2. Particle matching and tracking

The number of particles successfully segmented differed from image to image, mainly due to changes in lighting. To establish a set of reference particle shapes for tracking, all particles found in the first five images were concatenated. These reference particles were matched with subsequent images following a quasi-leapfrog updating scheme.

The distance calculated with Equation 5 between two descriptors is an arbitrary number. A threshold distance for a good match had to be established. The $FD_{dist}$ was calculated between all pairings of more than 500 particles in one of the reference images. The 2.5[th] percentile of the resulting distribution of distances was 0.13 and the 50[th] percentile was 0.25. Consequently, two particles in subsequent images were considered a match if the $FD_{dist}$ was less than 0.2. If the Fourier descriptor distance exceeded 0.15, the reference particle was updated to the median shape of the last five particles. In addition, a limit was set on the maximum allowable centroid displacement based on the applied strain (typically 0.1 mm per image). The Fourier descriptors of individual particles were smoothed across images to minimize local variations in shape.

If a particle was not found in the target image, a match was attempted in the next image in the sequence. The shape and position of the missing particle was linearly interpolated from the smoothed Fourier descriptors of the particle in the neighbouring images. Tracking of missing particles stopped once the median displacement of the remainder of the specimen reached 1.5 mm (approximately 50% of $D_{50}$).

Figure 11 shows the tracking of the vertical displacement of a particle across image frames for five cycles of loading. In addition, one of the two principal axes of the particle is shown. The principal axes were used to measure the rotation ($\Delta\theta$) of the particles. The principal axes of a shape are defined as the two axes coinciding with the minimum and maximum moments of inertia of the shape (Hearn 1997). To calculate the principal axes, the particles were represented as polygons with discrete vertices. These vertices were used to calculate the image moments following the method by (Steger 1996). Subsequently, the principal axes were calculated from the image moments (Mukundan and Ramakrishnan 1998).

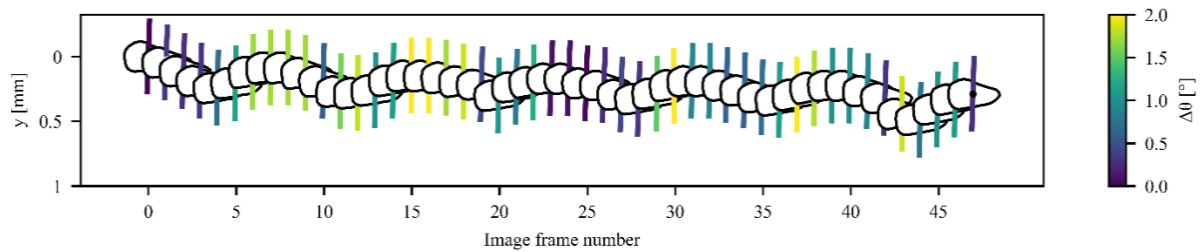

**Figure 11.** Tracking of the vertical displacement of and individual particle across image frames

### 4.3. Digital Image Correlation

A reference measurement of the sample's deformation was required to validate the measurements of the particle displacements. Digital Image Correlation (DIC) was applied to the images to measure these reference displacement fields. DIC is an optical technique used to measure full-field (continuum) displacements of a series of images (Schreier et al. 2009). Typically, the images are divided into a series of adjacent or overlapping subsets (or patches). The position of these subsets is tracked throughout a series of images by searching for the location in the target image where the cross-correlation with the reference subset is maximized. Sophisticated algorithms, such as the GeoPIV-RG software (Stanier et al. 2016) used in this study, allow for the subsets to deform and rotate during a test. The result of the DIC measurements is a field of displacement vectors across the image.

In the tests conducted for the present study, the fused quartz particles were comparatively large (~2.9 mm) in relation to the area of interest (~50 mm). This necessitated tracking relatively large subsets of ~10 mm in diameter to ensure that each subset contained at least three unique features (i.e. particles), as recommend by (Jones and Iadicola 2018). The overlapping subsets were spaced ~2.5 mm apart to provide a high spatial resolution of the measurements.

In contrast to the particle segmentation, the DIC analyses performed best on images that were not denoised with total variation denoising as well. This is likely due to total variation denoising removing some of the unique texture of the images in addition to the noise. Without this texture, the algorithm cannot uniquely match subsets.

When testing triaxial specimens with lubricated ends, the global displacement measurements represent the compression of both the soil and the lubrication (Sarsby et al. 1982). Consequently, local strain measurements are required for an accurate representation of the axial strain (Clayton and Khatrush 1986). The displacement fields from the DIC analyses were used to make local measurements of the axial strain. By considering the displacement vectors at the top and the bottom of the specimen, the average vertical strain of the specimen can be calculated. More details are provided in Marx and Zornberg (2022b).

### 4.4. Accuracy, resolution, and repeatability of the particle tracking

#### 4.4.1. Accuracy

The displacement fields generated using DIC were used to assess the accuracy of the particle displacement measurements. Figure 12 displays the horizontal and vertical displacement contours from the DIC analysis after 7.5% axial strain for one of the monotonic tests (CM-25-1). The segmented particles are overlaid on the DIC contours and shaded according to their displacement. For the particles shown in the figure, the median $FD_{dist}$ to the first image is 0.15. The relatively large $FD_{dist}$ can be attributed to inconsistencies in lighting and out-of-plane particle rotation at large strain.

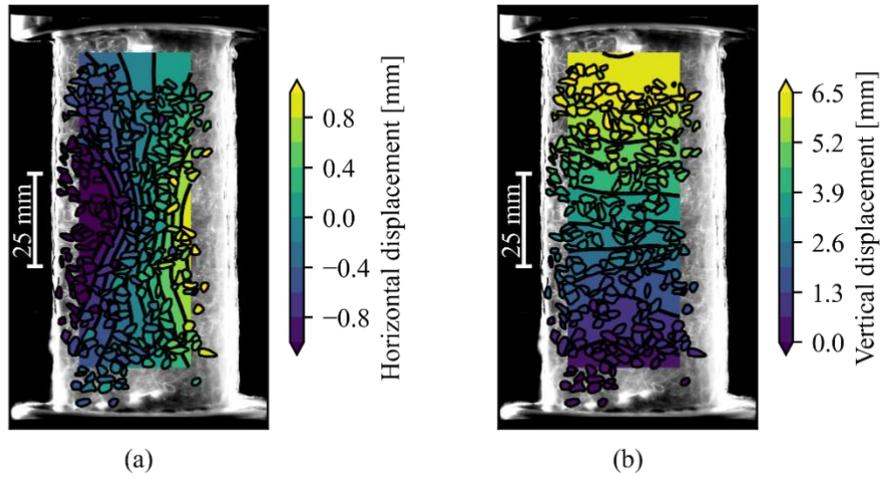

<div align="center">(a)         (b)</div>

**Figure 12.** Validation of particle displacement from segmentation and tracking against displacements measured with DIC for CM-25-1 at 7.5% axial strain: a) horizontal displacement; and b) vertical displacement.

The correlation coefficients between the displacements from the two methods at 7.5% axial strain were 0.87 and 0.99 for horizontal and vertical displacement, respectively. At 2.5% axial strain the correlation between the horizontal displacement measurements was only 0.6 while the corresponding correlation for vertical displacement was 0.96. At these levels of axial strain, shear bands had not yet formed, and localized horizontal particle displacement is haphazard. However, the large subsets of the DIC analysis smoothed these localized displacements, leading to poor correlation between the two measurement methods at low axial strain.

### 4.4.2. Resolution

The principal axis and centroid of a particle is sensitive to variations in its outline. To quantify this sensitivity (the resolution of the measurements), the outlines of 650 particles were randomly varied. Ten permutations were created for each particle. For each permutation the displacement of the centroid, the change in principal axis orientation and the distance between the new and original Fourier descriptors were calculated. The particles were reconstructed from the Fourier descriptors with 30 nodes. For each particle all the nodes were displaced in a random direction and magnitude. A uniform distribution was used to sample the random values.

The sensitivity of the particle orientations and centroid positions to outline are summarized in Table 3 for three median magnitudes of node offset. These sensitivities translate to the resolution of the method. For example, when the distance in Fourier descriptors between a reference and target particle was 0.06, the resolution of the principal angle measurements was 0.4° and the resolution of the centroid displacement measurements was 0.04 mm, i.e. approximately 0.04% axial strain.

Table 3. Sensitivity of particle orientation and centroid position to variations in particle outline.

| Median node offset [mm] | Median distance in Fourier descriptors [-] | Median, absolute change in orientation [°] | Median centroid displacement [mm] |
|---|---|---|---|
| 0.05 | 0.064 | 0.36 | 0.039 |
| 0.1 | 0.074 | 0.71 | 0.062 |
| 0.2 | 0.100 | 1.50 | 0.120 |

### *4.4.3.Repeatability*

The two monotonically sheared control specimens (CR-25-1 and CR-25-2) were used to confirm the repeatability of the specimen preparation procedure and the rotation measurements. Figure 13 shows the stress-strain response of the two specimens. Shearing of transparent sand is prone to so-called "slip-stick" behaviour. Slip-stick involves brittle drops in deviatoric stress, followed by a gradual increase in load, and is postulated to be due to chains of load-bearing particles forming, collapsing, and reforming within the specimen (Ezzein and Bathurst 2011). For the purpose of comparing the two repeat tests, these drops were filtered out of the stress-strain curves in Figure 13.

A good match was observed between the stress-strain responses of the two tests, confirming the repeatability of the test setup. In terms of volumetric behaviour, the samples matched up to ~7.5% axial strain. Both samples dilated significantly during shear as a result of the highly angular nature of the particles, the low confining stress adopted for the tests, as well as a high specimen density.

No strain softening of the deviatoric stress was observed, despite the high density and dilation of the specimens. This behaviour can be contributed to the specimens failing in diffuse bulging due to the use of lubricated platens (Hettler and Vardoulakis 1984). In addition, oil-saturated specimens tested by Ezzein and Bathurst (2011) and Kong et al. (2017) showed a lower peak strength than comparable water-saturated specimens despite having a similar residual strength.

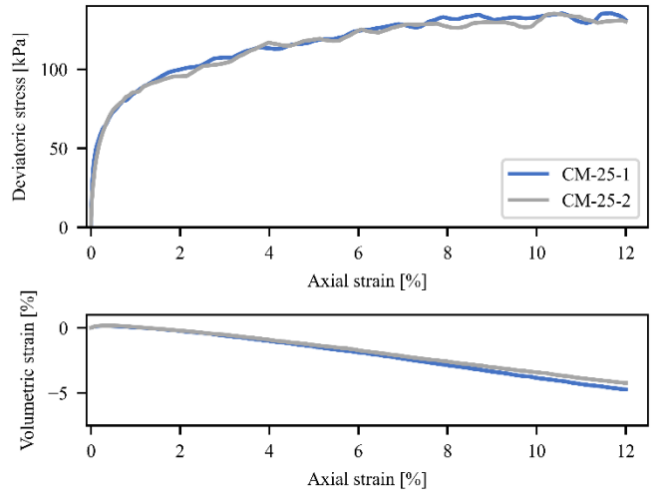

**Figure 13.** Stress-strain behaviour of the two monotonic tests.

The two monotonically sheared specimens were also used to investigate the repeatability of the rotation and displacement measurements. In Figure 14 the particle displacements and rotations at 7.5% axial strain on two planes through each test (East and West) are compared. The planes intersected at 60° to each other (see Figure 1). To minimize the impact of boundary effects, only particles in the centre 25 mm of the sample were considered. Across all four planes shown, the median $FD_{dist}$ to the reference images was less than 0.15.

Figure 14a shows that the median vertical particle displacement was similar across all the planes. The median, absolute horizontal displacement of the four planes shown in Figure 14b was similar with depth. However, there was more noise in the horizontal displacement measurement. The bulging that can be observed in the centre of the specimens, despite the use of lubricated platens, may be attributed to the testing of slender (2V:1H) specimens (Bishop and Green 1965).

In Figure 14c the median absolute rotation is compared with depth. There is larger variation in the rotation measured for the four planes. This variation is attributed to the lower resolution of the rotation measurements, as well as the greater inherent variation in particle rotation throughout a test. However, for all four planes the rotation was a maximum at the centre of the specimen and a minimum at the edges. Overall, the repeatability of the rotation and displacement measurements are deemed to be sufficient to investigate the mechanisms of soil-geogrid interaction.

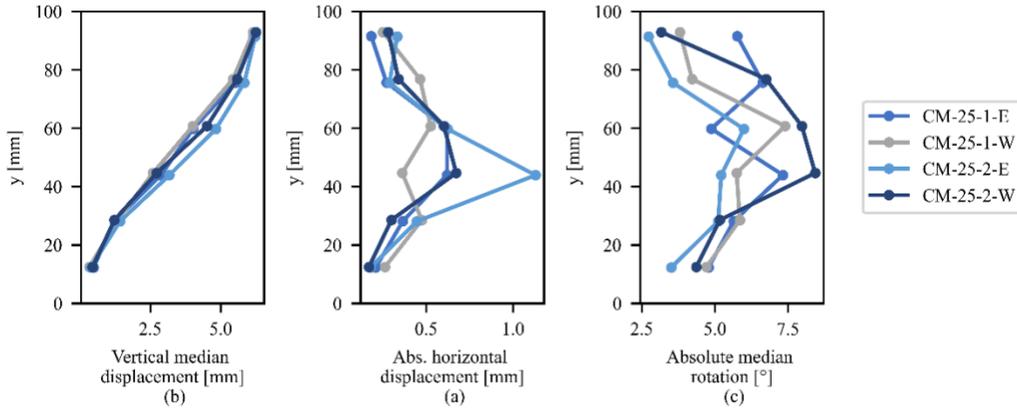

**Figure 14.** Comparison of particle motion with depth for the monotonic tests at 7.5% axial strain: (a) median vertical displacements; (b) median, absolute horizontal displacement and (c) median, absolute rotation.

### 4.5. State boundary line for particle motions

In Figure 15 the horizonal displacement ($\Delta x$) and absolute rotation ($|\Delta \theta|$) of the particles on a plane through H1-25 is shown with depth. Two stages of repeated loading (Stages 1 and 3), and the final stage of monotonic shear (Stage 4) are compared. The $FD_{dist}$ to the descriptors of the reference image are indicated by the colour bar.

Only the centre 75 mm of specimen was considered. In addition, particles near the membrane and those smaller than 1 mm in diameter were filtered from Figure 15. The median $FD_{dist}$ values for the three stages shown were 0.06, 0.08, 0.11 respectively. This is equivalent to a resolution in displacement and rotation measurements of (0.04 mm, 0.4°), (0.08 mm, 0.9°) and (0.12 mm, 1.5°) (refer to Table 3). This increase in the $FD_{dist}$ across the stages was due to the particle greater rotation and corresponding decreased match quality with vertical strain.

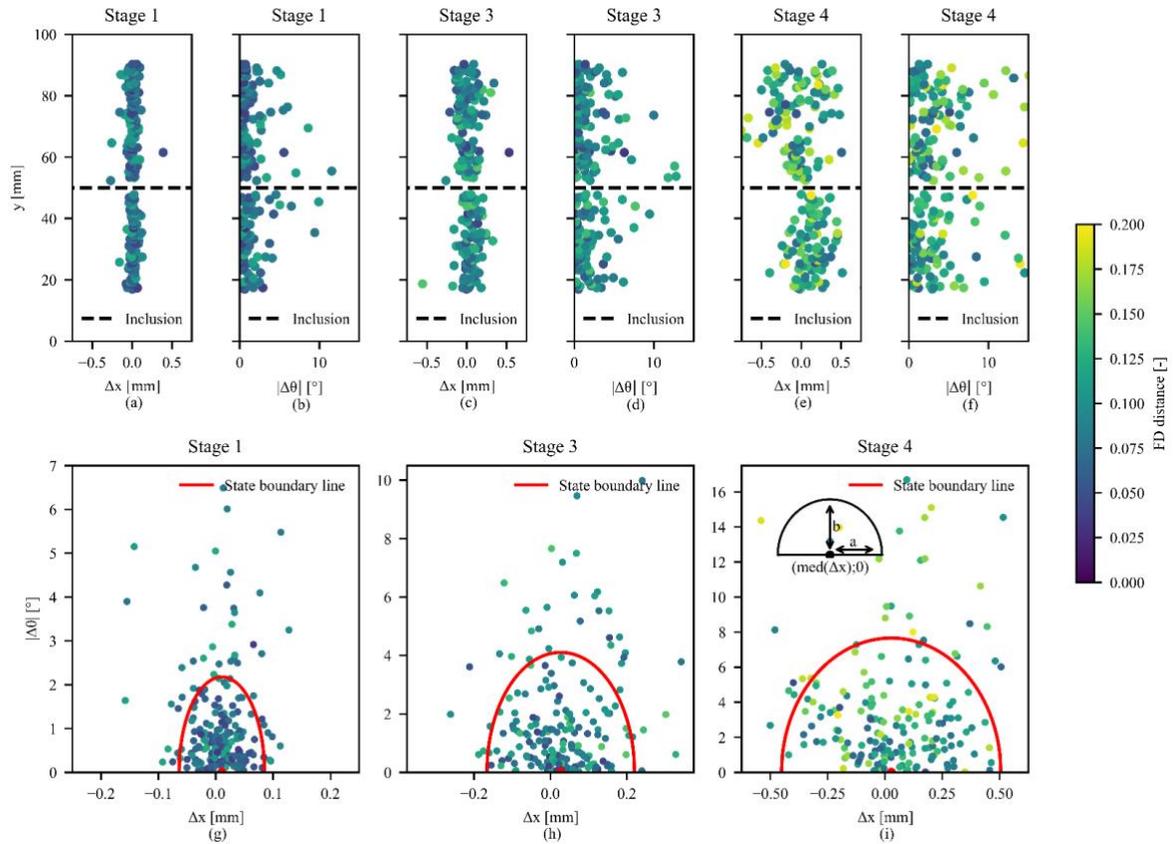

**Figure 15.** Horizontal displacement ($\Delta x$) and absolute rotation ($|\Delta\theta|$) of the segmented particles for Specimen H1-25 over three different stages of loading. (a) to (f): rotation and displacement as a function of depth. (g) to (i): rotation as a function of displacement. The points are coloured based on the $FD_{dist}$ of the particles to the reference particles.

In Stage 4 the local lateral restraint provided by the geosynthetic inclusion is apparent. Both the horizontal displacement (Figure 15e) and rotation (Figure 15f) of the particles reduced at the level of the inclusion. For Stages 2 and 3 (Figure 15a to d) the occurrence of lateral restraint is occluded by the presence of outliers.

Alternatively, the motion of each individual particle can be considered, i.e. the horizontal displacement – absolute rotation pair. In Figure 15g to 15i the particle motions are shown for three of the stages. These figures indicate that the particles motions tend to be weighted to either rotation or displacement. Large displacement combined with large rotation does not occur. Thus, a boundary line can be drawn separating probable states of displacement and rotation (motion) from improbable states of motion. Once the particle arrangement becomes unstable, e.g. at critical state, this boundary may not be relevant any more.

The state boundary lines (SBLs) in Figure 15g to 15i were drawn based on the assumption that all the particles in the centre 75 mm of the specimen are stiffened. This assumption follows the work of (McDowell et al. 2006) that found that the stiffened zone extended 100 mm from the geogrid (albeit for a $A/D_{50}$ of 6.84). From inspection of the data, the SBL was assumed to follow an ellipse centred on the x-axis:

$$\frac{(\Delta x - \Delta x_0)^2}{a^2} + \frac{|\Delta \theta|^2}{b^2} = 1 \tag{5}$$

where $\Delta x_0$ is the median horizontal displacement, and $a$ and $b$ are radii defined below.

To define the radii of the SBL outlier motions were identified using robust statistics. Outliers were defined as values for which the z-score was less than 2:

$$z = \frac{x_i - \text{median}(x)}{MAD(x)} < 2 \tag{5}$$

where $MAD(x)$ is the Median of Absolute Deviations from the median, a robust estimate of the variability of the data (Rousseeuw and Hubert 2018). Thus, the radii of the SBL were defined as:

$$a = 2 \cdot MAD(\Delta x) \tag{6}$$

$$b = \text{median}(|\Delta \theta|) + 2 \cdot MAD(|\Delta \theta|) \tag{7}$$

The extent of the SBL is sensitive to the selected $z$-value. Thus, the interpretation of the boundaries should be limited to establish trends rather than to obtain absolute values of motion limits.

## 5. Results

### 5.1. Global strain measurements

Figure 16 shows the accumulated, i.e. permanent, axial and radial strain after each unloading cycle for the four specimens tested under repeated loading. The axial strain was calculated from the DIC measurements. The radial strain is shown for an equivalent right cylinder deformed with the axial strain $\varepsilon_a$.

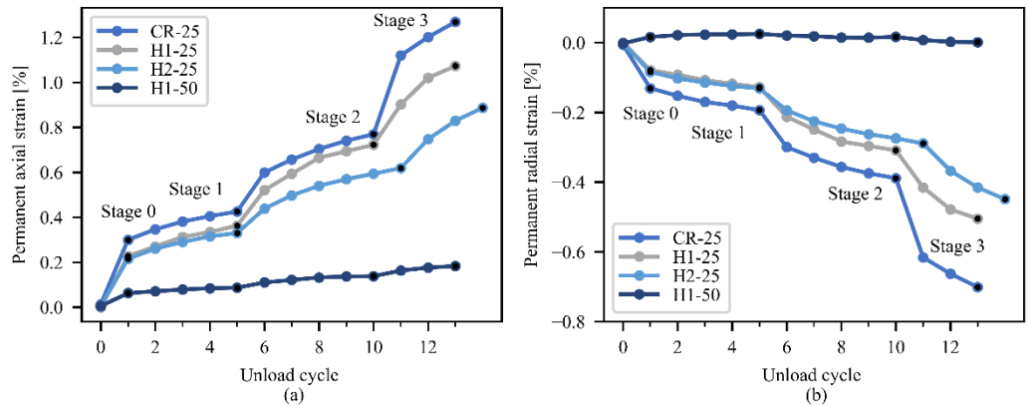

Figure 16. Accumulated strain in the specimens after each unload cycle: a) axial strain and b) radial strain. The end of each loading stage is also indicated.

A significant decrease in the rate of axial strain accumulation after the first unload cycle can be observed in Figure 16for all specimens. During the first loading cycles, particles typically rearrange into a more compact state. In railway ballast settlement is classified into a non-linear compaction phase ("shakedown") and a linear post-compaction (Jeffs and Marich 1987; Dahlberg 2001; Suiker et al. 2005). Thus, the first cycle of loading (Stage 0) was assumed to correspond to the "compaction" phase for the tests conducted in this study. The end of this load-unload cycle was considered the reference condition for subsequent calculations of rotation and displacement.

Lateral restraint provided by the inclusions H1 and H2 limited particle rearrangement, resulting in decreased permanent strain across all stages of loading. At the end of Stage 4 (monotonic shear until failure of the control specimen at $q = 140$ kPa) the axial strain was 8.7%, 3.15%, 2.8% and 0.5% for the CR-25, H1-25, H2-25 and H1-50 respectively. Of the two inclusions, H2 was the most effective in reducing permanent strain for the grading and particle size tested.

The interplay of dilation, contraction and lateral restraint by the geosynthetic inclusions confuses the interpretation of permanent volumetric strains for the stiffened specimens H1 and H2. A better understanding of the effect of the geosynthetic inclusions is gained from the radial strain shown in Figure 16b. For specimen H1-50, tested at the higher confining stress of 50 kPa, the radial strain is effectively zero. For the other specimens tested at 25 kPa confining stress, the trend in radial strain is also consistent with that of the permanent axial strains: H2 strains less than H1, which in turn strains less than the control specimen.

For the set of results generated in this study, a ratio of aperture size to $D_{50}$ of 2.2 (H2) proved to be more effective in stiffening the specimens than a $A/D_{50}$ ratio of 4 (H1). This value differs from the generally reported optimal $A/D_{50}$ ratio of 3.5 (e.g. Sarsby 1985). However, the uniformly graded, angular sand tested in this study is more similar to ballast than the soils tested (Sarsby 1985). For ballast tested with rectangular geogrids, an optimum $A/D_{50}$ in the order of 1 to 2 between aperture size and $D_{50}$ has been reported (Brown et al. 2007; Indraratna et al. 2013), which aligns with the results of this study.

### 5.2. Particle motion measurements

The SBLs for the four specimens tested under repeated load are compared in Figure 17 across the four stages of loading. The median $FD_{dist}$ to the reference images of the four tests is also indicated for each stage. For Stage 4, the SBL of CR-25 plotted outside the extends of the axis.

The base of a SBL represents the range of probable horizontal particle displacements. Across all four stages of loading, the geosynthetic inclusions restricted the probable horizontal displacements relative to the control specimen (CR-25). The degree of restriction was found to increase with increasing load. The geosynthetic inclusions also appeared to centre the column of sand, while the control specimen gradually shifted to the right.

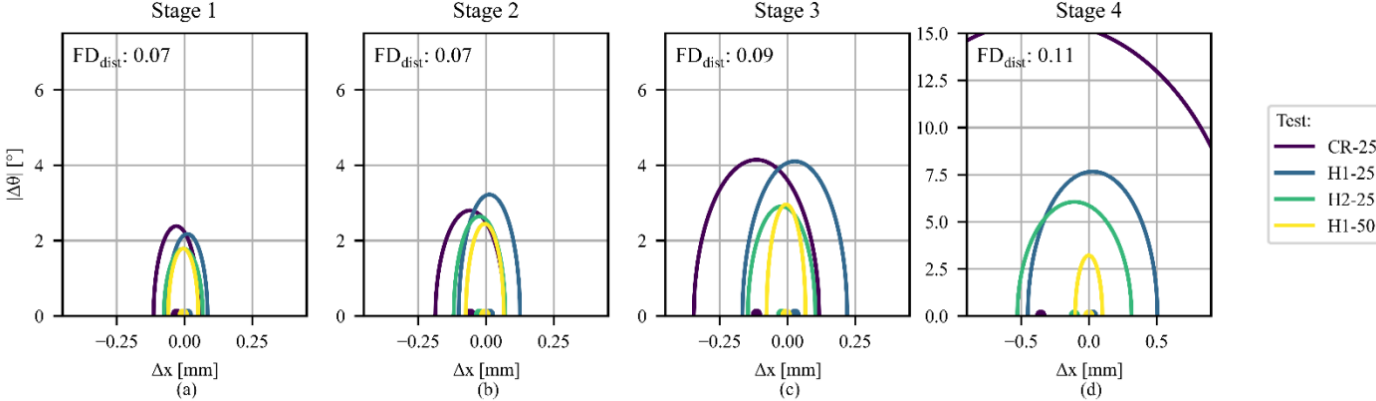

Figure 17. Comparison of the particle motion state boundary lines for the repeated load tests across four stages of loading: a) Stage 1, b) Stage 2, c) Stage 3 and d) Stage 4.

For the first two stages of loading, the difference in rotations measured among the specimens is finer than the resolution of the measurement technique (~0.4°). Thus, the limiting values of rotation cannot be reliably compared. Once the applied stress enters the non-linear regime in Stage 3 the difference in maximum probable rotation becomes significant. For inclusion H2, the maximum rotation of the particles was approximately 1° less than the control specimen. However, for specimen H1-25 the restriction in particle rotation by Stage 3 was still insignificant relative to the control specimen, despite the marked restraint in lateral particle displacement. Only at larger strains (Stage 4) did H1 restrict the particle rotations relative to the control specimen.

The results shown in Figure 17 illustrate that the combination of technologies involving transparent sand and deep learning-based segmentation can successfully be used to quantify the mechanism of lateral restraint provided by geogrids. The results in Figure 17 also show that the geosynthetic inclusions tended to restrict horizontal particle displacement more significantly than it restricted particle rotation. However, this may be due to the nature of triaxial tests, where particles tend to displace with applied load rather than rotate.

The area under the SBL represents the range of possible particle motions. In Figure 18 the permanent axial strain for the four specimens is compared to the area of the SBL. Both measurements made at the end of the unload stages and during monotonic shear are shown. The permanent strain shown during monotonic shear includes a negligible amount of elastic shear. The inset shows the measurements up to Stage 3 (i.e., the repeated load cycles).

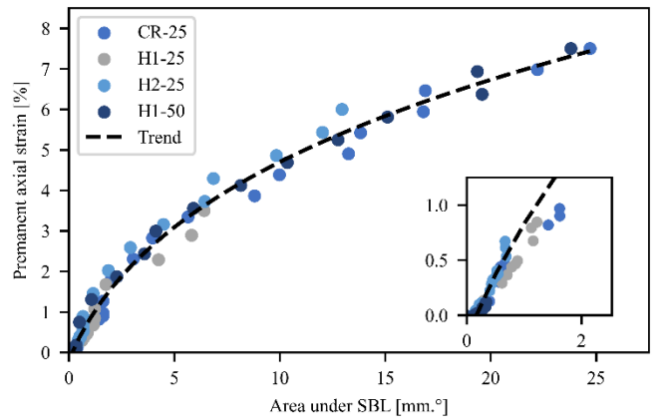

Figure 18. Permanent axial strain as a function of the area below the SBL for all four stages of loading. The inset shows the measurements up to Stage 3.

As the area under the SBL increased, the permanent strain of the specimens increased. For the set of data generated in this study, which involved a single soil density, and for the definition of the SBL as presented in Section 4.5, the correlation between the area under the SBL and permanent axial strain appears to be unique across confining stresses, loading stages and the presence and geometry of inclusions.

The unique relationship demonstrated in Figure 18 implies that the confining stress applied by a geogrid to an aggregate layer can now be quantified. Previously, the confining stress provided by a geogrid was predicted based on numerical studies (Kwon et al. 2008; Abu-Farsakh et al. 2014). Consider two hypothetical specimens that have both accumulated 0.1% permanent strain after 2500 load cycles. One is stiffened with a geogrid and was tested at 25 kPa confining stress, the other is a control specimen tested at 75 kPa confining stress. Both specimens are at the same permanent axial strain and will thus have the same range of probable particle motions. Consequently, the lateral restraint provided by the geogrid is equivalent to an additional confining stress of 50 kPa applied to the soil. In practice, control specimens might need to be tested at multiple confining stresses to interpolate the confining stress applied by the geogrid.

## 6.  Conclusions

This study illustrated the successful combination of deep learning-based segmentation and triaxial tests on transparent soil to quantify the lateral restraint provided by geogrids. The geogrids were simulated with geosynthetic inclusions. These inclusions were embedded in triaxial specimens subject to repeated loading. The Cellpose CNN was successfully trained to segment images of fused quartz. By considering the Score-CAM metric from the field of Explainable AI, it was shown that the CNN produced rational predictions, even though it was originally developed to segment biological cells. The results also showed that particles could be tracked between images despite minor variations in particle outlines by defining them in terms of their Fourier descriptors. The individual particle displacement measurements were found to correlate with displacement fields calculated using digital image correlation. While the resolution of the particle rotation measurements was somewhat coarser than the displacement measurements, repeatable measurements were still achieved.

By tracking the individual particle motions, a boundary between probable and improbable displacement-rotation states (motions) could be identified. The size of the zone of probable motions under a given applied load was found to decrease with increased lateral restraint, whether due to the presence of a geosynthetic inclusion or to an increased confining stress. The inclusions were found to reduce horizontal displacement of the particles more significantly than rotation. A unique relationship was found between the area under the state boundary line, i.e. the range of probable particle motions, and the permanent strain of the specimens. This unique relationship will allow for determination of the confining stress provided by a geogrid in an aggregate layer in future studies.

### Acknowledgement

The authors are grateful to Tensar International Corporation for the financial assistance provided for the study. Opinions expressed and conclusions presented are solely those of the authors.

### Appendix A1: Convolutional neural networks

A brief description of convolutional neural networks is provided as context for the discussion on the Cellpose architecture and Score-CAM method in Appendices A2 and A3. A common form of neural networks is a series of logistic regression models stacked together (Murphy 2012). For a single logistic model, the input layer $x$ is weighted with weights $v$, summed and passed through an activation function $g$, such as the logistic function. The final non-linear function, known as an activation function, distinguishes these models from standard linear regression.

Multiple logistic models can be applied to the same input. When the output of these multiple logistic models is fed into a subsequent logistic model, as shown in Figure 19, it is known as a neural network (multi-layer perceptron). The intermediate activations are known as hidden nodes ($z$) and forms a hidden layer. A network with multiple hidden layers is defined as a deep neural network (Goodfellow et al. 2016).

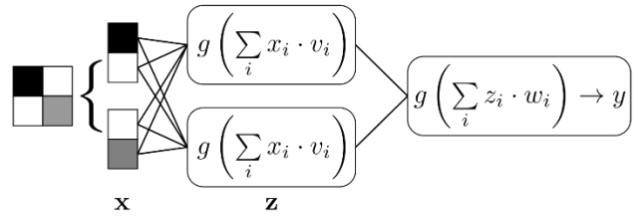

**Figure 19.** A multi-layer perceptron with two hidden nodes and a single output class.

In a fully connected network, such as the multi-layer perceptron in Figure 19, each input is connected to every node in the next layer. For image applications this is inefficient due to the large number of inputs (pixels). Consequently, a type of neural network known as a Convolutional Neural Network (CNN) is often used.

In a CNN the connections between pixels is sparse due to the use of convolution (Goodfellow et al. 2016). The process is demonstrated in Figure 20. An input image $x$ is convolved with a 3x3 filter $w_k$. That is, the cluster of nine input pixels is each multiplied with the corresponding weight of the filter. This result is summed to produce one entry in the output array. The process is repeated across the image, resulting in a filtered image.

The filtered (convolved) image is passed through a non-linear activation function ($g(\cdot)$), to calculate the feature map (or activation map) for that filter ($A_k$). Each convolutional layer typically consists of multiple filters that results in a series of feature maps. The feature maps in the final convolutional layer are combined into the output of the model, whether that is a denoised image, a segmented image or the probability of an object being present in the image.

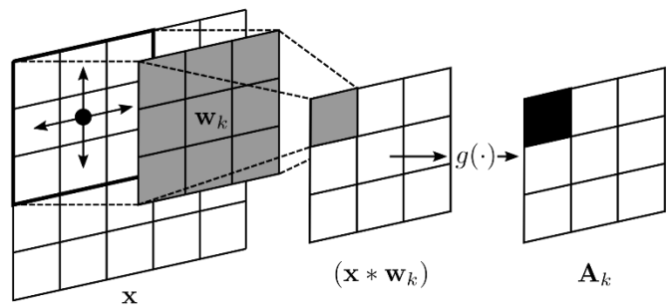

**Figure 20.** Convolution of an image with one filter

## Appendix A2: Cellpose architecture

Convolutional neural networks can be used for image segmentation. Cellpose implements a CNN architecture known as U-Net (Ronneberger et al. 2015). This architecture has high performance on limited training datasets at the cost of some general flexibility. The key components of U-Net are an encoder ($E$) followed by a decoder ($D$) as shown in Figure 21.

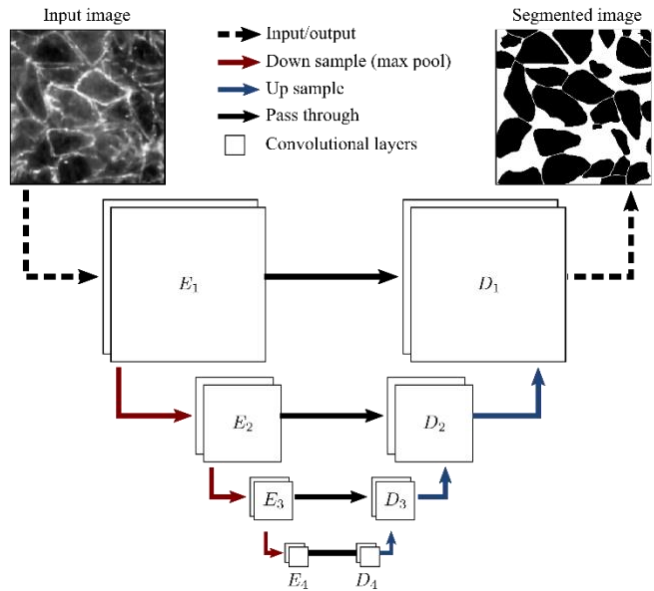

**Figure 21.** Key components of the U-Net architecture.

The network shown in Figure 21 consists of an encoder ($E_1$ to $E_4$) with four scales and decoder ($D_4$ to $D_1$) with four scales as well. Each scale consists of several, connected, convolutional layers with multiple scales. In the encoder the final activation maps from each scale are down-sampled using max pooling. The max pooling operation groups the feature maps into neighbourhoods of 2x2 pixels and then reports the maximum of each neighbourhood. These max pooled feature maps are used as input for the next scale. Typically, the number of features (filters) doubles between scales.

The last layer of the encoder ($E_4$) is connected to the first layer of the decoder ($D_4$) with convolution. In the decoder the input to each scale is the up-sampled output of the lower scale combined with the output of the equivalent level of the encoder. The feature maps of the final scale ($D_1$) are often passed to a fully connected layer to classify each pixel into a unique class. Subsequently, the image can be segmented based on the pixel classes.

Each entry of the feature maps of scale $E_4$ of the encoder represents a large region of pixels due to the down sampling between scales and convolution inside scales. Thus, any classification in the decoder that is derived from this value will be sensitive to the large region of pixels. This sensitivity is apparent in the Score-CAM heatmaps in Figure 9 where the pixel under consideration was affected by seemingly unrelated parts of the image.

The network used in Cellpose has several improvements on U-Net. Firstly, Cellpose uses residual layers (He et al. 2016), rather than standard convolutional layers. In addition, a vector representing the image 'style' or visual texture (see Gatys et al. (2016)), was calculated from the feature map of the smallest scale of the encoder. This style vector was added to the input of each of the scales of the decoder. By considering the image 'style' the network can more efficiently segment images of diverse styles such as fish scales, fused quartz and cancer cells.

The error, or loss, function for the network ($L_{CP}$) was defined as:

$$L_{CP} = \|\boldsymbol{y}_0 - 5\boldsymbol{H}\|^2 + \|\boldsymbol{y}_1 - 5\boldsymbol{V}\|^2 + L_{BCE}(\sigma(\boldsymbol{y}_2), \boldsymbol{P}) \tag{A1}$$

where $\boldsymbol{H}$ and $\boldsymbol{V}$ is the known horizontal and vertical gradients and $\boldsymbol{y}_0$ and $\boldsymbol{y}_1$ the corresponding prediction from the model. $\boldsymbol{P}$ is the known binary probability map, $L_{BCE}$ is the binary cross entropy loss function, $\sigma(x)$ the sigmoid function and $\boldsymbol{y}_2$ the predicted probability map.

The loss was minimized with stochastic gradient descent, using backpropagation to calculate the gradients. The version of the Cellpose model used in this paper was implemented in PyTorch (Paszke et al. 2019).

### Appendix A3: Score-CAM

The Score-CAM heatmap represents the contribution of each pixel in the image to the classification of a pixel under consideration. Each two-dimensional feature map in a CNN is activated by different characteristics of the image (e.g. ridges or empty space). Thus, different feature maps will be significant for the prediction depending on the class (e.g., particle or no particle) under consideration. The sum of all the feature maps, weighted by their significance to the class, results in a heatmap highlighting the regions of an image with the greatest contribution to the class under consideration.

For an image $X$, the Score-CAM heat map $M_{SC}$ for a class $c$ is defined as the sum of the $k$ feature maps $A^k$, of layer $l$, weighted by their contribution $\alpha_k^c$ to the class:

$$M_{SC}^c = ReLU\left(\sum_k \alpha_k^c A_l^k\right) \tag{A2}$$

where

$$\alpha_k^c = C\left(A_l^k\right) = f\left(X \circ H_l^k\right) - f(X_b) \tag{A3}$$

and

$$H_l^k = s\left(Up\left(A_l^k\right)\right) \tag{A4}$$

As presented by Equation A3, the contribution of a feature map ($\alpha_k^c$) to a class is defined as the difference between the class output for a baseline image ($f(X_b)$) and the output for an image where the features significant to class $c$ are emphasized $f\left(X \circ H_l^k\right)$. $H_l^k$ defined in Equation 5 is the feature map up-sampled ($Up(\cdot)$) to the size of the image ($X$) and normalized between 0 and 1 ($s(\cdot)$). Finally, the weighted sum is passed through the Rectified Linear Unit operator (ReLU, $f(x) = \max(0, x)$) as only positive contributions to a class are relevant.

For U-Net the feature maps of the convolutional layers of the encoder are typically the most insightful to analyse (Vinogradova et al. 2020). Thus, the heatmaps shown in Figure 9 is for the final convolutional layer of the encoder.